\documentclass[10pt,conference]{IEEEtran}
\IEEEoverridecommandlockouts
\usepackage{cite}
\usepackage{amsmath,amssymb,amsfonts}
\usepackage{algorithmic}
\usepackage{graphicx}
\usepackage{textcomp}
\usepackage{xcolor}
\usepackage{color}
\usepackage{float}
\usepackage{subfig}
\usepackage{overpic} 
\usepackage{makecell}
\usepackage{multirow}

\def\BibTeX{{\rm B\kern-.05em{\sc i\kern-.025em b}\kern-.08em
    T\kern-.1667em\lower.7ex\hbox{E}\kern-.125emX}}
\begin{document}

\title{{\sc ACWRecommender}: A Tool for Validating Actionable Warnings with Weak Supervision}

\author{
\IEEEauthorblockN{Zhipeng Xue\IEEEauthorrefmark{2},
Zhipeng Gao\thanks{* corresponding author.}\IEEEauthorrefmark{1}\IEEEauthorrefmark{2},
Xing Hu\IEEEauthorrefmark{2},
Shanping Li\IEEEauthorrefmark{2}
} \\
\IEEEauthorblockA{
\IEEEauthorrefmark{2}Zhejiang University, Hangzhou, China}

\IEEEauthorblockA{
\{ zhipengxue, zhipeng.gao, xinghu, shan\}@zju.edu.cn
}
}

\maketitle

\begin{abstract}
Static analysis tools have gained popularity among developers for finding potential bugs, but their widespread adoption is hindered by the accomnpanying high false alarm rates (up to 90\%). 
To address this challenge, previous studies proposed the concept of actionable warnings, and apply machine-learning methods to distinguish actionable warnings from false alarms. 
Despite these efforts, our preliminary study suggests that the current methods used to collect actionable warnings are rather shaky and unreliable, resulting in a large proportion of invalid actionable warnings. 
In this work, we mined 68,274 reversions from Top-500 Github C repositories to create a substantia actionable warning dataset and assigned weak labels to each warning's likelihood of being a real bug. 
To automatically identify actionable warnings and recommend those with a high probability of being real bugs (AWHB), we propose a two-stage framework called {\sc \textbf{ACWRecommender}}.
In the first stage, our tool use a pre-trained model, i.e., UniXcoder, to identify actionable warnings from a huge number of SA tool's reported warnings. 
In the second stage, we rerank valid actionable warnings to the top by using weakly supervised learning. 
Experimental results showed that our tool outperformed several baselines for actionable warning detection (in terms of F1-score) and performed better for AWHB recommendation (in terms of nDCG and MRR).  
Additionaly, we also performed an in-the-wild evaluation, we manually validated 24 warnings out of 2,197 reported warnings on 10 randomly selected projects, 22 of which were confirmed by developers as real bugs, demonstrating the practical usage of our tool.
\end{abstract}


\section{Introduction}
The static analysis tool has been widely used by software developers and companies to detect potential bugs and report warnings in recent years~\cite{1559604, 7476667, 10.1145/2970276.2970347}.
For example, Facebook has developed \texttt{infer}, a static code analysis tool for checking generic bug patterns (e.g, null pointer exceptions, memory leaks, race conditions) in their Android and iOS apps (including the main Facebook, Whatsapp, Instagram app and many others). 
Due to the lightweight analysis and low computational cost, these static analysis tools have gained popularity among developers. 
However, static analysis tools face two major challenges: Firstly, they have a high false alarm rate, with warnings often having a false-positive rate of reaching up to 90\%~\cite{HECKMAN2011363}. 
Secondly, developers often become overwhelmed with information overload while using these tools, which can cause them to overlook real bugs and getting lost in irrelevant information.

To fill the gap and help developers to better make use of static analysis tools, previous researchers have introduced the concept of \textbf{actionable warning}, namely the warnings need to be acted on by developers.~\cite{Alikhashashneh2018UsingML, Heo2017MachineLearningGuidedSU}. 
Particularly, if a warning presents in one revision and disappears in a subsequent revision, then this warning is referred as an actionable warning, otherwise, it is regarded as a false alarm. 
Different methods have been proposed to identify actionable warnings~\cite{wang2018there, Yuksel2013AutomatedCO, Lee2019ClassifyingFP}. 
However, two limitations still exist: 
(i) Most of these methods rely on hand-crafted features, which heavily depend on manual design and expert domain knowledge. 
(ii) Prior research has focused solely on detecting actionable warnings without addressing the crucial concern that not all actionable warnings are valid indicators of real bugs.  

In this paper, we aim to automate the task of identifying actionable warnings produced by static analysis tool (\texttt{infer} in this study) and further validatin actionable warnings by recommending \textbf{AWHB} (\textbf{A}ctionable \textbf{W}arning with \textbf{H}igh probability to be real \textbf{B}ug). 
To achieve this, we build a large dataset of actionable warnings and propose a two-stage framework called {\sc ACWRecommender} which includes a coarse-grained detection stage and a fine-grained reranking stage. 
The dataset is collected from the Top 500 popular C projects on Github, consisting of 538 actionable warnings and 30,590 false alarms. 
Then each actionable warning is assigned a weak label using semantic and structural matching rules to estimate the likelihood of it being a real bug. 
In the coarse-grained detection stage, we train a detector to predict whether a warning is actionable or not using the dataset. In the fine-grained reranking stage, we fine-tune the model to prioritize AWHB using weakly supervised learning.

We conducted extensive experiments on two tasks, actionable warnings detection and AWHB recommendation, to evaluate the effectiveness of our {\sc ACWRecommender}. Our proposed model demonstrated superiority over several baselines. Results show that our first-stage detector significantly outperforms other baselines by 91.7\% in terms of F1-score for the actionable warning detection task, and our second-stage reranker performs better than its three baselines in terms of nDCG and MRR for the AWHB recommendation task. An in-the-wild evaluation was also conducted, where our tool recommended 24 actionable warnings to Github developers, 22 of which were confirmed as real bugs, further justifying the practical usage of our approach.

\label{sec:intro}

\section{Approach}
\label{sec:approach}
\begin{figure*}[htbp]
	\centering
    \centerline{\includegraphics[width=0.99\textwidth]{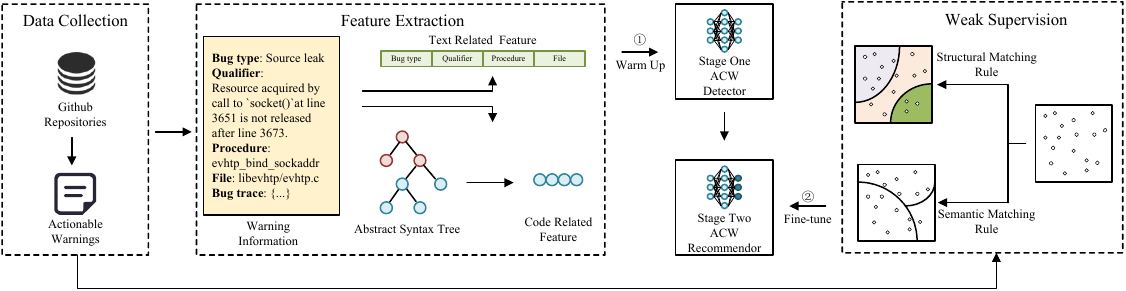}}
    \caption{Overview of Our Tool}
    \label{fig:overview}
\end{figure*}

We first introduce how to build an actionable warning dataset under weak supervision. We then present the details of our proposed two-stage model. The Overall framework is illustrated in Fig.~\ref{fig:overview}

\subsection{Actionable Warning Collection and Labeling}
The aim of this stage is to gather all actionable warnings and assign a label for each actionable warning to represent its likelihood of being a real bug. 
Regarding the actionable warning collection, we followed the process of previous studies~\cite{wang2018there}, 
for each revision a given project, we run static analysis tool \texttt{infer} to generate a list of warnings. 
Then we automatically check if the warning disappears in later revisions, if yes then the warning is labeled as an actionable warning, otherwise the warning is treated as a false alarm. 
Any warnings that have not been resolved for over two years are also deemed false alarms. 
Finally, we have collected 538 actionable warnings and 30,590 false alarms from top-500 GitHub C projects. 
However, we find that \textbf{actionable warnings collected by the current pipeline are largely invalid and may not necessarily represent real bugs}, this observation is also consistent with the latest empirical findings~\cite{kang2022detecting}. 
The main reason is that the current method regards all disappeared warnings as actionable warnings, while this assumption is rather shaky because the disappearance of such warning(s) can be caused by a non-relevant fix. 
In this study, we aim to take one more step further by assigning actionable warnings with different probability scores under weak supervision. 
The higher probability scores indicate actionable warning(s) are more likely to be real bug(s) (referred to \textbf{AWHB} in this study), which should be inspected at the beginning. 
To gather more accurate actionable warnings, we estimate the ``matching degree'' of each actionable warning and its bug-fix commit in terms of two perspectives, i.e., semantical matching rule (using commit message) and structural matching rule (using code change context). 

\subsubsection{Semantic Matching Rule.}
The commit message of a bug-fix reversion summarizes the commit and can be used to estimate a semantic matching score. 
As shown in Table~\ref{Table:Semantic Matching Rule}, a score of 3 is assigned if the commit message contains keywords that exactly match the warning type (column 3), indicating a high likelihood of a real bug. 
A score of 2 is assigned if warning context keywords are mentioned in the commit message (column 4). 
A score of 1 is assigned if the commit message only contains fix-related common keywords (column 5). 
If the commit message does not match any keyword, a score of 0 is assigned, indicating an unlikely relation to a real bug.


\begin{table*}[htbp]
  \begin{center}
    \caption{Semantic Matching Rule}
    \label{Table:Semantic Matching Rule}
    \begin{tabular}{|c|c|c|c|c|} 
      \hline
        Warning Type & Warning Qualifier Template & Warning Type Keyword & Warning Context Keyword& Common Keyword\\
        \hline
        \hline
        Uninitialized Variable& \makecell{The value read from \textit{variable} \\was never initialized.} & initial, define, assign, etc.& \textit{variable} &\multirow{4}*{\makecell[c]{\\\\fix, repair, bug, \\warning, solve, \\problem, etc.}}\\
        \cline{0-3}
        Null Dereference&\makecell{\textit{pointer} last assigned on line \# could \\be null and is dereferenced at line \# }& \makecell{null dereference,\\null pointer, etc.}&\textit{pointer} &\\
        \cline{0-3}
        Resource Leak &\makecell{Resource acquired to \textit{variable} \\by call to \textit{function} at line \# \\is not released after line \#} &resource, leak, etc. &\textit{variable}, \textit{function} &\\
        \cline{0-3}
        Dead Store& \makecell{The value written to\\ \textit{variable} is never used.} & \makecell{dead store, \\unused, never, etc.}& \textit{variable}&\\
      \hline
    \end{tabular}
  \end{center}

\end{table*}

\subsubsection{Structural Matching Rule.}

Besides semantic information, we use code change context to assist determination if an actionable warning is fixed by the corresponding bug-fix reversion. 
As shown in Table~\ref{Table:Structural Matching Rule}, a structural matching score is assigned based on code change matching rules. A score of 3 is assigned if the code change matches the fix pattern of the warning type (column 2). 
A score of 1 is assigned if the code change falls within expected bug-fix scope (column 3). For warnings that do not match any rule, a score of 0 is assigned.

\begin{table}[htbp]
  \begin{center}
    \caption{Structural Matching Rule}
    \label{Table:Structural Matching Rule}
    \begin{tabular}{|c|c|c|} 
  \hline
    Warning Type & Fix Pattern & Scope Pattern \\
    \hline
  \hline
    Uninitialized Variable & \makecell{assign value by \\assignment or reference} & before warning\\
    \hline
    Null Dereference & add a null-check & before warning\\
    \hline
    Resource Leak & \makecell{invoke resource-\\free-related function} & after warning\\
    \hline
    Dead Store & \makecell{use the variable,\\remove assignment} & after warning\\
      \hline
    \end{tabular}
  \end{center}
     \vspace{-10pt}
\end{table}

\subsubsection{Aggregation of Weal Labels}
To obtain a more reliable and robust label for each actionable warning, 
we use majority voting to combine the above semantic matching score and structural matching score, as demonstrated in Equ.~\ref{Equ:Label Aggregation Eqution}.

\begin{equation}
  \begin{gathered}
\label{Equ:Label Aggregation Eqution}
  Label(x) = \begin{cases}
    {\bf \texttt{VTB}}, & CM(x)+CC(x)>3 \\
    {\bf \texttt{LTB}}, & 2<=CM(x)+CC(x)<=3 \\
    {\bf \texttt{UTB}}, & 0<=CM(x)+CC(x)<2 \\
  \end{cases}
  \end{gathered}
\end{equation}

Given an actionable warning $x$, $CM(x)$ refers to the semantic score for $x$ using the commit message matching rule, and $CC(x)$ refers to the structural score for $x$ using the code change matching rule. 
If the sum of $CM(x)$ and $CC(x)$ is greater than 3, it means the actionable warning is very likely to be a real bug from both semantic and structural aspects and we label $x$ as {\bf \texttt{VTB} (Very Likely To be Bugs)}. 
Similarly, if the sum of $CM(x)$ and $CC(x)$ falls between 2 and 3, it means the warning $x$ matches the bug-fix reversion from either the semantic or structural aspect and is likely to be a real bug, we label $x$ as {\bf \texttt{LTB} (Likely To be Bugs)}. 
Lastly, if the sum of $CM(x)$ and $CC(x)$ is less than 2, it means the actionable warning $x$ mismatches the bug-fix reversion and is unlikely to be a real bug, we label such instances as {\bf \texttt{UTB} (Unlikely To be Bugs)}. 
We then define the warning $x$ whose $Label(x)$ is {\bf \texttt{VTB}} or {\bf \texttt{LTB}} as \textbf{AWHB} (\textbf{A}ctionable \textbf{W}arning with \textbf{H}igh probability to be real \textbf{B}ug). 

\subsection{Two-stage Model}
In order to identify actionable warnings and recommend AWHB, we propose a two-stage modal which includes a detector and a reranker.
In the coarse-grained detection stage, actionable warnings dataset is used to warm-up and let the model learn how to distinguish actionable warnings from false alarms. 
In the fine-grained reranking stage, we further fine-tune the model to rerank the AWHB to the top by weakly supervised learning.

\subsubsection{Warn-up the detector model}
The goal of this stage is to develop a model that can differentiate between actionable warnings and false alarms. To achieve this, UniXcoder~\cite{Guo2022UniXcoderUC} is used to process two types of key information: text-related input (such as bug type) and code-related input (such as AST). The use of UniXcoder allows for unified multi-modal data training, which is ideal for the encoding task.
The text-related input is obtained by extracting bug type, qualifier, procedure, and filename from the warning report generated by \textit{infer}. The code-related input is generated by identifying the bug location and associated buggy statement, locating the parent node of the statement in the AST, and extracting control flow information from the AST as code context. Both types of input are fed into UniXcoder, and the resulting embeddings are obtained by concatenating the pre-trained model outputs. The model is trained to identify actionable warnings by warming up UniXcoder.
The actionable warning identification task can be viewed as a binary classification problem. 
That is, for a given reported warning $x$, we use the model $f(x ; \boldsymbol{\theta})$ to determine whether $x$ is actionable or not. 
The actionable warning dataset without weak supervision is used to warm up $f(x ; \boldsymbol{\theta})$ and the optimization goal is defined as follows:

\begin{equation}
  \begin{gathered}
\min _{\boldsymbol{\theta}} \frac{1}{N} \sum_{x \in \mathcal{X}} \mathcal{L}\left(f(x ; \boldsymbol{\theta}), {y}_{x}\right)
  \end{gathered}
\end{equation}

where $\mathcal{L}$ denotes a loss function and ${y}_{x}$ is the actionable warning label without weak supervision. Any loss function suitable for a classification task can be used in the warm-up process, and in this study we use the Binary Cross Entropy Loss.

\subsubsection{fine-tune the reranker model}
Based on the warmed-up actionable warning detector model, we fine-tuned a reranking model by continuing training with the actionable warning dataset under weak supervision.
To rank different levels of actionable warnings, we transform the ranking problem into multiclass classification task.
That is, for a given reported warning $x$, we aim to predict $x$ as \texttt{V}/\texttt{L}/\texttt{U}/\texttt{False Warning} based on our actionable warnings dataset under weak supervision. 
The optimization problem is defined as follows:

\begin{equation}
  \begin{gathered}
\min _{\boldsymbol{\theta}} \frac{1}{N} \sum_{x \in \mathcal{X}} \mathcal{J}\left(g(f(x ; \boldsymbol{\theta})), \tilde{y}_{x}\right)
  \end{gathered}
\end{equation}

The $\tilde{y}_{x}$ is the label aggregated from weak supervision. $g(x)$ is the softmax function to compute the probability of each class for $x$, and $\mathcal{J}$ is Cross Entropy Loss, which is suitable for multiclass classification task. 
When the optimization is done, the final ranking score for each warning $x$ can be inferred as follows:

\begin{equation}
  \begin{gathered}
\label{Equ:Ranking Score}
  \mathcal{S}(x) = \begin{cases}
    class(x) + g_{\tilde{y}}(x), & \overline{y} \in \texttt{VTB}, \texttt{LTB}, \texttt{UTB} \\
    class(x) - g_{\tilde{y}}(x), & \overline{y} \in \texttt{False Warning} \\
  \end{cases}
  \end{gathered}
\end{equation}


where $class(x)$ maps each warning $x$ to a base class score 0/1/2/3 if the predicted class is \texttt{False Warning}/\texttt{UTB}/\texttt{LTB}/\texttt{VTB}, $\overline{y}$ denotes the predicted class of $x$ and $g_{\tilde{y}}(x)$ denotes the probability of the predicted class. 

\subsection{Implementation}
We implement our tool by Python 3.9, leveraging the PyQt5~\cite{Summerfield2007RapidGP}, which is a popular Python GUI module. 
To use our tool, developers first input a \texttt{infer} report (txt file from their local computer) by selecting the local file and clicking the \texttt{open} button, then the developers click the \texttt{Analyze} button in the bottom of the tool interface to call our pre-trained model by feeding the input infer reports. 
In the backend, {\sc ACWRecommender} can read each warning from the report, pre-process the necessary text-related and code-related information automatically, then {\sc ACWRecommender} identifies the potential actionable warnings and rerank AWHB to the top. 
Finally, the top returned warnings will be listed in the right side of our tool, we also highlight the warnings with red and orange colours to help developers to address the AWHB with high priority.

\begin{figure}[htbp]
	\centering
    \centerline{\includegraphics[width=0.48\textwidth]{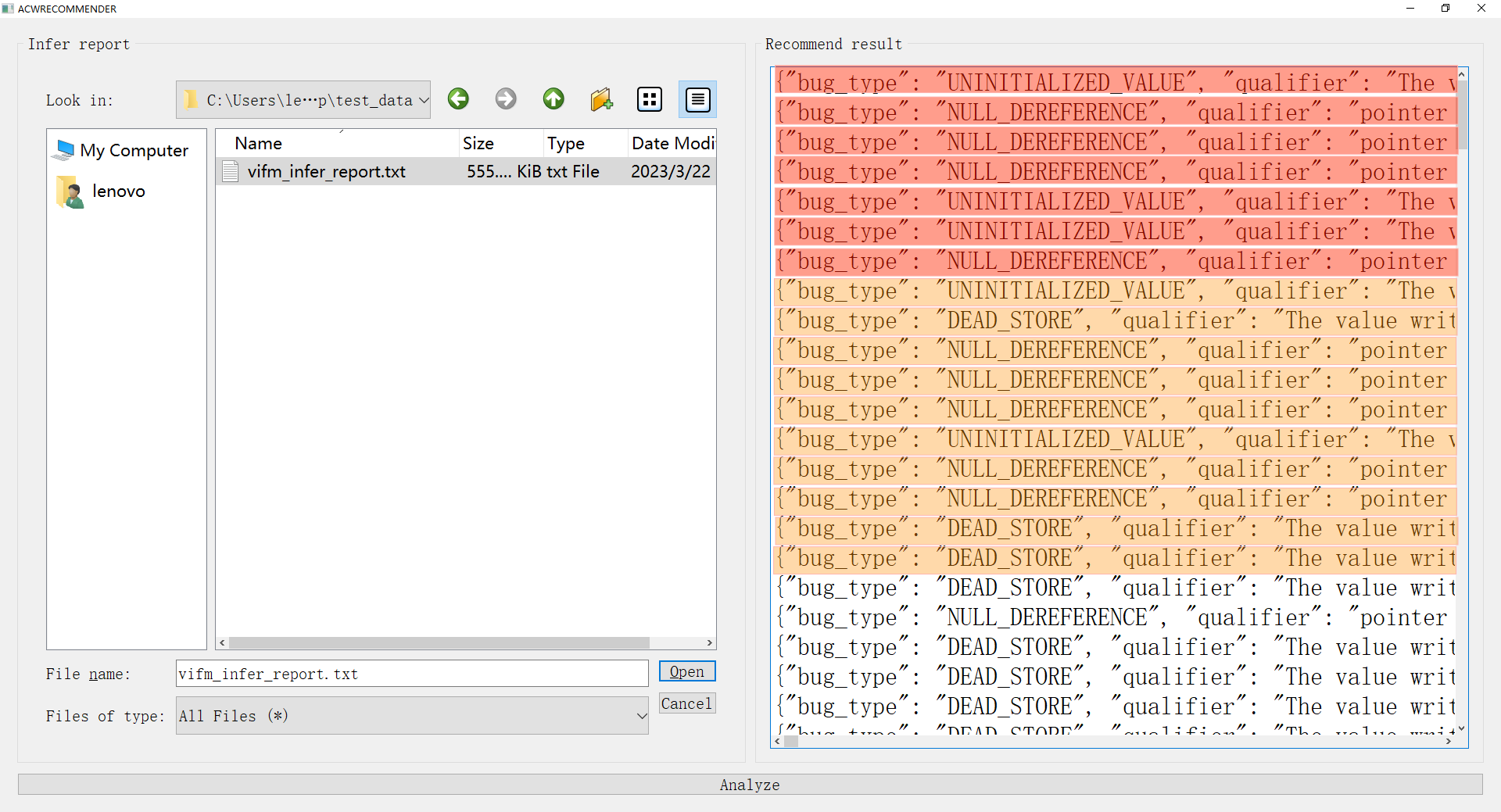}}
    \caption{User Interface of {\sc ACWRecommender}}
    \label{fig:tool}
\end{figure}

\section{Evaluation}
\label{sec:eval}

To evaluate the performance of our tool, we first build actionable warning dataset from the top 500 repositories (ordered by the number of stars) in Github for C repositories. In this study, we use \textit{infer} with the default settings to generate warnings, which primarily fall into four categories: uninitialized variable, resource leak, null dereference, and dead store. Finally, we collect both actionable and false warnings from 68,274 revisions of the 394 projects that we reminded, including 538 actionable warnings and 30,590 false warnings. According to our weak supervision rules. 
Among the 538 actionable warnings, we assign the 57 of them as {\bf \texttt{VTB}}, 59 {\bf \texttt{LTB}} and 422 {\bf \texttt{UTB}}. We only regard the actionable warning whose aggregated labels are {\bf \texttt{VTB}} and {\bf \texttt{LTB}} as \textbf{AWHB}.

Then we compare {\sc ACWRecommender} with three baselines: Golden Feature~\cite{wang2018there}, Random Forest and Random Selection/Ranking. Golden Feature is a stats-of-the-art tool for actionable warning identification, which uses the manually defined golden features (e.g., added lines, number of methods) and leveraged the SVM model. 
Random Forest uses a random forest model which performed the best in our preliminary experiments, to identify and recommend actionable warning based warning feature embedding. 
Random Selection/Ranking is a popular baseline, which identifies the actionable warning based on the ratio of actionable warning datset.

 \subsection{Quantitative Analysis}
 \subsubsection{RQ1: The Identification Effectiveness Evaluation}
To evaluate the effectiveness of our proposed tool, i.e., {\sc ACWDetector}, we evaluate it and the baseline methods on our testing set in terms of Precision, Recall and F1-score. The evaluation results is shown in Table~\ref{Table:Identification Effectiveness Evaluation}.

\begin{table}[htbp]
  \begin{center}
    \caption{Identification Effectiveness Evaluation}
    \label{Table:Identification Effectiveness Evaluation}
    \begin{tabular}{|c|c|c|c|} 
    \hline
    Measure & Precision & Recall & F1-score\\
    \hline
    \hline
    \textbf{Random Selecting} & 0.019 & 0.019 & 0.019 \\
    \hline
        \textbf{Golden Feature} & 0.294 & 0.183 & 0.154 \\
    \hline
    \textbf{Random Forest} & 0.433 & 0.217 & 0.289 \\
    \hline
{\sc \textbf{ACWDetector}} & \textbf{0.472} & \textbf{0.671} & \textbf{0.554} \\
    \hline
    \end{tabular}
  \end{center}
\end{table}

\textbf{Our proposed tool, {\sc ACWDetector}, surpasses all the baseline methods by a significant margin in terms of all evaluation metrics.} In specific, {\sc ACWDetector} increase the Precision, Recall and F1-score of Random Forest 9.0\%, 184.3\% and 91.7\%, respectively.

\subsubsection{RQ2: The Recommendation Effectiveness Evaluation}
Since the AWHB should be recommended and handled earlier, we evaluate the recommendation effectiveness of our reranker tool. We build 100 queries by randomly selecting 1,000 warnings from testing set. We set K to 1, 3, and 5 for nDCG@K metric. The evaluation result is listed in Table~\ref{Table:Recommendation Effectiveness Evaluation}.

\begin{table}[htbp]
  \begin{center}
    \caption{Recommendation Effectiveness Evaluation}
    \label{Table:Recommendation Effectiveness Evaluation}
    \resizebox{0.5\textwidth}{11mm}{
    \begin{tabular}{|c|c|c|c|c|} 
    \hline
    Measure & nDCG@1 & nDCG@3 & nDCG@5 & MRR \\
    \hline
    \hline
    \textbf{Random Ranking} & 0.020 & 0.031 & 0.052 & 0.007\\
    \hline
        \textbf{Golden Feature} & 0.074 & 0.095 & 0.137 & 0.064\\
    \hline
    \textbf{Random Forest} & 0.143 & 0.211 & 0.239 & 0.204 \\
    \hline
{\sc \textbf{ACWRecommonder}} & \textbf{0.212} & \textbf{0.381} & \textbf{0.416}  &\textbf{0.396}\\
    \hline
    \end{tabular}}
  \end{center}
\end{table}


 \textbf{Our proposed tool is effective for AWHB recommendation, and outperforms all the baseline methods.} The nDCG@1 value exceeds 0.2, indicating that over one-fifth of the queries can accurately identify AWHB in the topmost position of the recommended warning list with a high degree of certainty. The MRR value is approximately 0.4, suggesting that on average, the first AWHB can be found at the third position in the recommended warning list. 

\subsection{In-the-wild Evaluation}

Our goal is to help developers find real bugs from static analysis tools while inspecting as few warnings as possible. 
We conducted an in-the-wild evaluation to assess the practical value of our {\sc ACWRecommender} in real-world Github repositories. We randomly selected 10 C repositories from GitHub with more than 500 stars and ran \textit{infer} and obtained 2,197 reported warnings. {\sc ACWRecommendor} reranked the warnings and generated a list for checking. We manually checked 659 (Top 30\%) warnings and found 24 potential bugs, of which 22 were confirmed by developers, with 20 merged and 2 approved by code reviewers. Notably, \textbf{12 of the 22 confirmed bugs were identified within the Top 10\% of recommended warnings}, demonstrating the effectiveness of our tool in reducing the effort required to find real bugs among a large number of warnings.

\section{Conclusion}
\label{sec:conclusion}
This research aims to identify actionable warnings produced by static analysis tool and recommend the Actionable Warning with High probability to be real Bug.  To address these task, we first collects actionable warnings from top-500 Github repositories. To the best of our knowledge, this is the first actionable warning dataset collected by mining the histories of popular Github repositories. We propose an approach named {\sc ACWRecommender}, which leverages UniXcoder to learn the semantic features of warnings and involves a coarse-grained detection stage and a fine-grained reranking stage. Extensive experiments on the real-world Github repositories have demonstrated its effectiveness and promising performance.

\section*{ACKNOWLEDGMENT}
\label{sec:ack}
This research is supported by the Starry Night Science Fund of Zhejiang University Shanghai Institute for Advanced Study, Grant No. SN-ZJU-SIAS-001. 
This research is partially supported by the Shanghai Rising-Star Program (23YF1446900) and the National Science Foundation of China (No. 62202341). 

\bibliographystyle{IEEEtran}
\bibliography{reference}

\end{document}